# Continuous and discrete Schrödinger systems with PT-symmetric nonlinearities

Amarendra K. Sarma[1,*], Mohammad-Ali Miri[2], Ziad H. Musslimani[3], and Demetrios N. Christodoulides[2]

[1]*Department of Physics, Indian Institute of Technology Guwahati, Guwahati-781039, Assam, India*
[2]*CREOL/College of Optics, University of Central Florida, Orlando, Florida 32816, USA*
[3]*Department of Mathematics, Florida State University, Tallahassee, Florida 32306-4510*
[*]aksarma@iitg.ernet.in

We investigate the dynamical behavior of continuous and discrete Schrödinger systems exhibiting parity-time (PT) invariant nonlinearities. We show that such equations behave in a fundamentally different fashion than their nonlinear Schrödinger counterparts. In particular, the PT-symmetric nonlinear Schrödinger equation can simultaneously support both bright and dark soliton solutions. In addition, we study a discretized version of this PT nonlinear Schrödinger equation on a lattice. When only two-elements are involved, by obtaining the underlying invariants, we show that this system is fully integrable and we identify the PT-symmetry breaking conditions. This arrangement is unique in the sense that the exceptional points are fully dictated by the nonlinearity itself.

## I. Introduction

Interest in non-Hermitian systems has been on the rise since the pioneering work of Bender and Boettcher where they showed that a wide class of non-Hermitian Hamiltonians can exhibit entirely real spectra as long as they respect the conditions of parity and time (PT) symmetry [1]. Since then PT-symmetric Hamiltonian families have been a subject of intense research within the context of quantum mechanics [2-6]. In general, a necessary condition for a Hamiltonian $H = -\frac{1}{2}\frac{d^2}{dx^2} + V(x)$ to be PT-symmetric is that the complex potential satisfies $V^*(-x) = V(x)$. Under this condition, the spectrum of the Schrödinger equation $H\psi = E\psi$ can be completely real. This is of course true as long as the system resides in the exact phase regime. On the other hand, if the imaginary component of this potential exceeds a certain threshold, the so called PT-symmetry breaking threshold, the PT symmetry will spontaneously break down and the spectrum will cease to be entirely real.

Recently, it has been suggested that optics can provide an ideal test bed for observing and studying the ramifications of such theories. This is due to the fact that, in optics, the paraxial equation of diffraction is mathematically isomorphic to the Schrödinger equation in quantum mechanics [7-9]. This analogy allowed observation of PT-symmetry in optical waveguide structures and lattices [8-10]. In addition, several studies have showed that PT symmetric optics can lead to new classes of optical structures and devices with altogether new properties and functionalities [7-23]. These include power unfolding and breaking of the left-right symmetry [7], abrupt phase transitions [8], non-Hermitian Bloch oscillations [12], simultaneous lasing-absorbing [13,16], and selective lasing [18]. Moreover, unidirectional invisibility [23] and defect states with unconventional properties [10,23] have been also demonstrated. Finally PT-symmetric concepts have also been used in plasmonics[22], optical metamaterials [23] and coherent atomic medium [24].

On the other hand, nonlinear Schrödinger systems involving PT symmetric linear potentials $(i\psi_z + \frac{1}{2}\psi_{xx} + V(x)\psi + |\psi|^2\psi = 0)$ have been intensely investigated within the last few years [25-35]. For example, these works include the effect of nonlinearity on beam dynamics in parity-time symmetric potentials [25], solitons in dual-core waveguides [26,27], nonlinear suppression of time reversal [28], dynamics of a chain of interacting PT-invariant nonlinear dimers [29], Bragg solitons in nonlinear PT-symmetric periodic potentials [30], and nonlinear interactions in PT-symmetric oligomers [31,32].

In general however, the Kerr nonlinearity can dynamically induce an effective linear potential which may not necessarily be PT-symmetric. As a result, this effective potential can dynamically break the even symmetry required for the real part of a PT potential. Once this symmetry is lost, the wave evolution in such nonlinear system is no longer bounded and hence a PT-breaking instability can ensue. Lately, Ablowitz and Musslimani considered for the first time the integrability of a new class of nonlinear highly non-local Schrödinger-like equations [36]. In this equation the standard third order nonlinearity $|\psi|^2\psi$ is replaced with its PT-symmetric counterpart $\psi(x,z)\psi^*(-x,z)\psi(x,z)$. Interestingly, in this study it was shown that this equation is fully integrable since it possesses linear Lax pairs and an infinite number of conserved quantities [36].

In this work we study the PT-symmetric Nonlinear Schrödinger Equation (PTNLSE) in continuous media as well as in discrete systems. Our analysis indicates that the PTNLSE exhibits altogether new behavior in terms of solutions and dynamics. In particular, this equation admits both bright and dark solitons under the exact same conditions. In addition we study a discretized version of this equation in an infinite lattice of coupled elements and then in a two-element coupled system. We show that such PT coupler is fully integrable. This article is structured as follows. In Sec. II the PT-symmetric nonlinear Schrodinger equation is presented. Afterwards, in Sec. III we present a discretized version of the PT-symmetric nonlinear Schrodinger equation in an array of coupled elements. In Sec. IV we consider the two-element PT-symmetric nonlinear coupled system, followed by a stability analysis in Sec. V. The integrability of the PT-symmetric nonlinear coupler is discussed in Sec. VI followed by conclusions in Sec. VII.

## II. PT-symmetric nonlinear Schrödinger equation

The PT symmetric nonlinear Schrödinger (PTNLSE) can be obtained from the standard nonlinear Schrödinger equation after replacing $\psi^*(x,z)$ with $\psi^*(-x,z)$. In other words:

$$i\psi_z + \frac{1}{2}\psi_{xx} + \psi(x,z)\psi^*(-x,z)\psi(x,z) = 0 \quad (1)$$

This equation can in principle be viewed as a linear Schrödinger-like equation $i\psi_z + \frac{1}{2}\psi_{xx} + V(x,z)\psi(x,z) = 0$ with a *self-induced potential* of the form $V(x,z) = \psi(x,z)\psi^*(-x,z)$. Such a dynamic potential is parity-time symmetric in the sense that $V(x,z) = V^*(-x,z)$. It should be noted that Eq. (1) is nonlocal, i.e. the evolution of the field at the transverse coordinate $x$ always requires information from the opposite point $-x$. These type of nonlinearities may be found in various wave mixing phenomena under appropriate PT-symmetric settings. In passing, it should be noted that nonlocal nonlinearities are ubiquitous in nature, for example, it may arise from the fluctuation of the external linear potential confining the wave, as in the case of BEC's in spatially and temporally fluctuating trapping potentials, diffusion of charge carriers or atoms/molecules in atomic vapors [37,38]. Nonlinearities are also nonlocal in case of optical beams in nonlinear dielectric waveguides or waveguide arrays with random variation of refractive index, size, or waveguide spacing [39]. In addition, long-range interactions of molecules in nematic liquid crystals also result in nonlocal nonlinearities [40]. We emphasize that Eq. (1) describes a non-hermitian system. In fact by defining the total power, $P = \int_{-\infty}^{+\infty} dx |\psi|^2$, one can directly show that power is not conserved during evolution and $\frac{dP}{dz} = \int_{-\infty}^{+\infty} dx |\psi|^2 [\psi\psi^*(-x,z) - \psi^*\psi(-x,z)]$.

Before going into details, it is worth noting that, in direct analogy with the standard Schrödinger equation, one can find an infinite number of constant of motions for Eq. (1) [36]. Here we mention the quasi-power, $Q$, and the Hamiltonian, $H$ of this system [36]:

$$Q = \int_{-\infty}^{+\infty} dx \psi(x,z)\psi^*(-x,z) \quad (2a)$$

$$H = \frac{1}{2}\int_{-\infty}^{+\infty} dx [\psi_x(x,z)\psi_x^*(-x,z) - \psi^2(x,z)\psi^{*2}(-x,z)] \quad (2b),$$

where $\psi_x$ represents the first derivative of $\psi$ with respect to $x$. These quantities can be obtained from their NLSE counterparts simply by replacing $\psi^*(x,z)$ with $\psi^*(-x,z)$.

It is straightforward to show that the PTNLSE, Eq. (1), admits a bright soliton solution:

$$\psi(x,z) = A\,\text{sech}(Ax)\exp\left(i\frac{A^2}{2}z\right) \quad (3)$$

Interestingly, unlike the standard NLSE, the PTNLSE of Eq.(1) admits at the same time a dark soliton solution as well:

$$\psi(x,z) = A\tanh(Ax)\exp(-iA^2 z) \quad (4),$$

where $A$ in both cases is a real constant representing the amplitude of these soliton states. Note that the standard NLSE can support only one of these two solutions, depending on the sign of dispersion or that of the nonlinear term. Furthermore one can show that the PTNLSE admits any symmetric solution of the NLSE (having positive nonlinearity) as well as any anti-symmetric solution of the NLSE with negative nonlinearity. This may include higher-order soliton solutions [41] as well as travelling soliton waves provided they are taken in symmetrically positioned pairs.

It should be noted however that, in stark contrast with the standard NLSE, the solutions of PTNLSE are not invariant with respect to shifts in the transverse coordinate $x$. In fact the solutions of Eq. (1) retain their shape during evolution in $z$ as long as it remains centered around the origin of the $x$ coordinate. For any shift from the center, the self-induced potential $V(x,z) = \psi(x,z)\psi^*(-x,z)$ nonlinearly breaks its PT symmetry in spite of the fact that it always respects the necessary condition of PT-symmetry, i.e. $V^*(-x,z) = V(x,z)$. This spontaneous breaking of PT-symmetry could be explained as follows: at a reference propagation distance $z_0$, if $\psi(x,z_0)$ is symmetric or anti-symmetric in $x$, i.e. $\psi(-x,z_0) = \pm\psi(x,z_0)$, then the dynamic potential is completely real $V(x,z_0) = \pm|\psi(x,z_0)|^2$. If on the other hand, the field distribution is asymmetric, the dynamic potential exhibits an imaginary part which is necessarily anti-symmetric. As long as this anti-symmetric imaginary part is below a certain threshold, the system is stable. However when the imaginary part increases above the threshold, this local PT-symmetry spontaneously breaks down and results in exponential growth of the field, triggering instability.

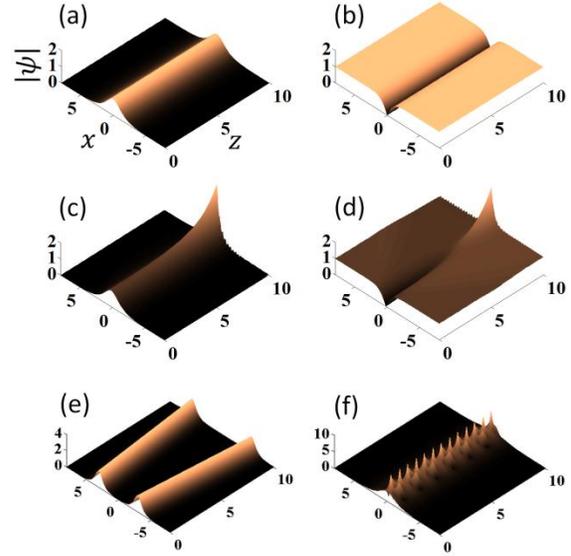

FIG. 1 (Color online) Numerical simulations of the propagation of the bright and dark solitons: (a) Bright soliton (b) Dark soliton (c, d) bright and dark solitons solutions becomes unstable when shifted from the center of PT symmetry, (e) pair of symmetrically positioned traveling solitons, (f) higher order soliton.

The evolution dynamics of Eq. (1) when initially excited with either the bright or dark soliton solutions of Eqs.(3,4) are depicted in Fig. 1(a) and 1(b) respectively. As it was expected, such solutions retain their shape during propagation in $z$. On the other hand, Figures 1(c) and 1(d) show the evolution of these same solutions when slightly shifted from the center. Clearly, such solutions are unstable as a result of spontaneous PT symmetry breaking. The possibility of travelling soliton pairs and higher-order PTNLSE solitons is also depicted in Figs. 1 (e,f).

It is not clear how this new type of nonlinearity could be realized practically. However, taking cues from various recent experimental results related to PT-symmetry in optics[8,9], it seems that coupled waveguide system or infinite array of waveguides may finally help us to realize such nonlinearities. Clearly, it may be worthwhile to study a discrete version of the PTNLSE, embodying the nonlinearity of Eq. (1).

## III. Discrete PT-symmetric nonlinear Schrodinger equation

To better understand the PT-symmetric nonlinear term in continuous systems, perhaps it is beneficial to see how it plays a role in discrete settings. In this section, we consider a discrete version of PTNLSE. This can be done by discretizing the transverse coordinate $x$ into the discrete lattice sites; $= 0, \pm 1, \pm 2, \ldots$ . Under this conditions the discrete PTNLSE can be written as:

$$i\frac{da_n(z)}{dz} + \kappa(a_{n+1}(z) + a_{n-1}(z)) + \rho a^*_{-n}(z)a_n^2(z) = 0 \quad (5)$$

where $a_n$ denotes the field amplitude at the discrete lattice site $n$, $\kappa$ represents the linear coupling coefficient between adjacent sites and $\rho$ is the coefficient of the PT-symmetric nonlinearity. According to Eq. (5) the field at location $n$ is linearly coupled to adjacent sites $n-1$ and $n+1$ while it is nonlinearly coupled to the field at mirror site $-n$. Inspired by the invariant parameters of the continuous PTNLSE, it is straightforward to show that the discrete PTNLSE admits the following constant of motions:

$$Q = \sum_n a_n a^*_{-n} \quad (6)$$

$$H = -\sum_n \left[\kappa(a_n a^*_{-n+1} + a^*_{-n} a_{n+1}) + \frac{\rho}{2}\left(a_n^2 {a^*_{-n}}^2\right)\right]. \quad (7)$$

Stationary soliton solutions of Eq. (5) can be found by assuming $a_n = A_n \exp(i\mu z)$ which leads to $\kappa(A_{n+1} + A_{n-1}) + \rho A^*_{-n} A_n^2 = \mu A$. Obviously, by assuming even $(A_{-n} = A_n)$ and odd $(A_{-n} = -A_n)$ solutions, this last equation turns into the discrete NLSE in a standard array of optical waveguide with focusing and defocusing nonlinearity [42]. As a result, Eq. (5) admits standard solutions of nonlinear waveguide arrays with both focusing and defocusing nonlinearity at the same time. Numerical results (based on Newton-Raphson method) show the presence of all such solitary wave solutions. In general however, such discrete solitons lack an analytical expression. On the other hand, as we will show in the next section, a two-element coupler with a PT-symmetric nonlinearity is fully integrable.

## IV. PT-symmetric nonlinear coupler

We next consider the discrete PTNLSE where only two elements are taken into account. In this regard we study a two-dimensional system embodying the nonlinearity of Eq. (1). This is expressed by a system of two coupled differential equations describing a PT-symmetric nonlinear coupler:

$$i\frac{da(z)}{dz} + \kappa b(z) + \rho b^*(z)a^2(z) = 0 \quad (8a)$$

$$i\frac{db(z)}{dz} + \kappa a(z) + \rho a^*(z)b^2(z) = 0 \quad (8b)$$

This set of coupled equations describes the physical situation reasonably well when a CW beam is launched into a array of two waveguides, each waveguide exhibiting a nonlocal PT-symmetric nonlinearity as described above. Here $a$ and $b$ represent the modal field amplitudes in the nonlinear coupler, $\kappa$ is the coupling constant and $\rho$ is associated with the strength of the nonlinearity. As opposed to standard nonlinear Kerr couplers [43,44], here the nonlinearity obeys PT symmetry. It is important to note that this arrangement is PT symmetric in a nonlinear sense as opposed to other systems where this symmetry is introduced in a linear fashion [35]. As in the continuous case, these coupled equations describe a non-conservative system. In other words, unlike a Hermitian system, the total power in the system $P = |a|^2 + |b|^2$ is not conserved. On the other hand, it is straightforward to show that the discrete counterparts of the quasi-power and Hamiltonian invariants do exist and are given by:

$$Q = a^*b + b^*a \quad (9)$$

$$H = -\kappa(|a|^2 + |b|^2) - \frac{\rho}{2}\left(a^2 b^{*2} + b^2 a^{*2}\right) \quad (10)$$

The presence of these two constants of the motion implies that Eqs. (8) are in fact integrable. In direct analogy with the standard nonlinear coupler [45] one can find the following two nonlinear supermodes of Eqs. (8):

$$\begin{pmatrix}a\\b\end{pmatrix} = \begin{pmatrix}+A_0\\+A_0\end{pmatrix} e^{i(+\kappa + \rho A_0^2)z} \quad (11a)$$

$$\begin{pmatrix}a\\b\end{pmatrix} = \begin{pmatrix}+A_0\\-A_0\end{pmatrix} e^{i(-\kappa - \rho A_0^2)z} \quad (11b)$$

Here the parameter $A_0$ is an arbitrary real constant. It should be noted that the first and the second solutions represent the symmetric and anti-symmetric nonlinear supermodes of the standard nonlinear coupler [45] in the presence of focusing and defocusing nonlinearity respectively. Quite interestingly, one can show that in addition to these one parameter supermodes, Eqs. (8) also admit a pair of *fixed point* nonlinear supermodes

$$\begin{pmatrix}a\\b\end{pmatrix} = \sqrt{\kappa/\rho}\begin{pmatrix}+1\\+e^{+i\theta}\end{pmatrix}e^{+i2\kappa\cos(\theta)z} \quad (12a)$$

$$\begin{pmatrix}a\\b\end{pmatrix} = \sqrt{\kappa/\rho}\begin{pmatrix}+1\\-e^{-i\theta}\end{pmatrix}e^{-i2\kappa\cos(\theta)z} \quad (12b)$$

In this case there is a phase difference of $\theta$ between the two channels. Even though the amplitudes are fixed to $\sqrt{\kappa/\rho}$ the phase difference $\theta$ can take any arbitrary values between 0 to $2\pi$. Specifically, for $\theta = 0$ these two solutions reduce to the symmetric and anti-symmetric solutions of Eqs. (8) with $A_0 = \sqrt{\kappa/\rho}$. Again, for $\theta = \pi/2$ these two solutions collapse to the stationary solution $(a, b) = \sqrt{\kappa/\rho}\,(1, i)$. It is worth noting that, in general the fixed point solutions appear in dissipative nonlinear systems involved with gain and loss [46]. As it was mentioned before, here the nonlinearity solely plays the role of an effective gain or loss in this system.

## V. Stability analysis of the supermodes

The stability of the nonlinear supermodes of Eqs. (8) can be investigated in the same manner as in continuous media. To investigate the stability of the supermode of Eq. (11a), we consider the following solution:

$$\begin{pmatrix}a\\b\end{pmatrix} = \begin{pmatrix}+A_0 + \varepsilon(z)\\+A_0 - \varepsilon(z)\end{pmatrix}e^{i(+\kappa+\rho A_0^2)z} \quad (13)$$

where $|\varepsilon| \ll A_0$ is a small perturbation to the supermode. Using this relation in Eqs. (8) and after neglecting higher order terms in $\varepsilon$ one finds the following evolution equation for the perturbation $\varepsilon$:

$$i\frac{d\varepsilon}{dz} - 2\kappa\varepsilon + \rho A_0^2\big(\varepsilon(z) - \varepsilon^*(z)\big) = 0 \quad (14)$$

After using the ansatz $\varepsilon = \varepsilon_r \cos(\mu z) + i\varepsilon_i \sin(\mu z)$ in Eq. (14) and solving the underlying eigenvalue problem one gets $\mu^2 = 4\kappa^2\left(1 - \frac{\rho}{\kappa}A_0^2\right)$. Therefore the even supermode (Eq.(11a)) is stable as long as $-\sqrt{\kappa/\rho} < A_0 < +\sqrt{\kappa/\rho}$. Similarly one can study the stability of the anti-symmetric supermode (Eq.11(b)) under the action of perturbations as follows:

$$\begin{pmatrix}a\\b\end{pmatrix} = \begin{pmatrix}+A_0 + \varepsilon(z)\\-A_0 + \varepsilon(z)\end{pmatrix}e^{i(-\kappa-\rho A_0^2)z} \quad (15)$$

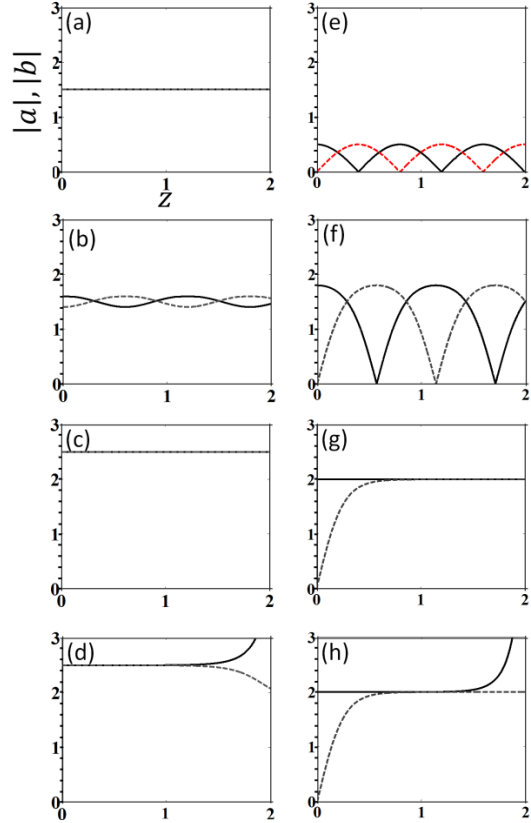

FIG. 2. Evolution dynamics of $a$ (black line) and $b$ (dashed gray line) for different initial conditions when $\kappa = 4$ and $\rho = 1$. (a) $a_0 = b_0 = 1.5$, (b) $a_0 = 1.6, b_0 = 1.4$, (c) $a_0 = b_0 = 2.5$, (d) $a_0 = 2.50001, b_0 = 1.49999$, (e) $a_0 = 0.5, b_0 = 0$, (f) $a_0 = 1.8, b_0 = 0$, (g) $a_0 = 2, b_0 = 0$, (h) $a_0 = 2.000001, b_0 = 0$.

which leads to the exact same stability region, i.e., the odd supermode is also stable as long as $-\sqrt{\kappa/\rho} < A_0 < +\sqrt{\kappa/\rho}$. Finally, one can show that the fixed point solutions are always unstable.

A numerical study of Eqs. (8) justifies these results. Figure (2) depicts the evolution dynamics of $a$ and $b$ for different initial values $a(z = 0) = a_0$ and $b(z = 0) = b_0$. The evolution of the symmetric supermode with and without perturbations is plotted in Fig. 2(a-d) in two different regimes $A_0 < \sqrt{\kappa/\rho}$ and $A_0 > \sqrt{\kappa/\rho}$. On the other hand, Fig. 2(e-h) depict the evolution dynamics for a single channel excitation where $b_0 = 0$.

According to this figure, Eqs. (8) exhibit a threshold-like behavior that resembles the linear PT-symmetric coupler [18]. Indeed, as we will show in the next section, $\sqrt{\kappa/\rho}$ is a critical value showing the onset of PT-symmetry breaking. If the initial values of any of these two variables $a$ and $b$ exceed $\sqrt{\kappa/\rho}$, the system becomes unstable.

## VI. PT-symmetric nonlinear coupler: Stokes parameters-analysis

In this section, by using the Stokes parameters of the system we further investigate the integrability of this PT-symmetric nonlinear coupler. We define the set of Stokes parameters as follows:

$S_0 = aa^* + bb^*$ (16a)
$S_1 = aa^* - bb^*$ (16b)
$S_2 = a^*b + b^*a$ (16c)
$S_3 = i(a^*b - b^*a)$ (16d)

It should be noted that $S_2$ is a constant of the motion (see Eq. (9)). By definition all these parameters are real and satisfy the following condition:

$S_0^2 = S_1^2 + S_2^2 + S_3^2$ (17)

By using Eqs. (16) and (17) it is straightforward to show that the evolution of the Stokes parameters is governed by the following set of nonlinear equations:

$\frac{dS_0}{dz} = -\rho S_1 S_3$ (18a)
$\frac{dS_1}{dz} = 2\kappa S_3 - \rho S_0 S_3$ (18b)
$\frac{dS_2}{dz} = 0$ (18c)
$\frac{dS_3}{dz} = -2\kappa S_1$ (18d)

Equations (18a) and (18d) together lead to $\frac{dS_0}{dz} = \frac{\rho}{4\kappa} \frac{dS_3^2}{dz}$ which shows that $S_0 = \frac{\rho}{4\kappa} S_3^2 + C$ where $C = S_0 - \frac{\rho}{4\kappa} S_3^2$ is another constant of motion. In addition, Eqs. (18b) and Eqs. (18d) lead to $\frac{d^2 S_3}{dz^2} = -2\kappa(2\kappa S_3 - \rho S_0 S_3)$. Combining these two latter relations we reach to the following equation:

$\frac{d^2 S_3}{dz^2} = \frac{\rho^2}{2} S_3^3 + (2\kappa\rho C - 4\kappa^2)S_3$ (19)

This is the so called Duffing equation which can be solved analytically by using Jacobian elliptic functions [47]. Instead of solving this equation however, here we restrict our attention in finding the PT instability criterion. As we will see this can also be obtained by simple graphical methods. In order to find the onset of this PT-symmetry breaking instability, we first assume that the coupler is excited with the initial condition $a(z=0) = a_0$ and $b(z=0) = b_0$ which are in general complex. From Eqs. (17) and (18), we have:

$S_0 = \frac{\rho}{4\kappa} S_3^2 + C$ (20a)
$S_0^2 = S_1^2 + S_3^2 + D$ (20b)

Here the two constants $C = S_0 - \frac{\rho}{4\kappa} S_3^2$ and $D = S_2^2$ can be uniquely determined in terms of the initial conditions:

$D = a_0^{*2} b_0^2 + a_0^2 b_0^{*2} + 2|a_0|^2 |b_0|^2$ (21a)
$C = |a_0|^2 + |b_0|^2 + \frac{\rho}{4\kappa}(a_0^{*2} b_0^2 + a_0^2 b_0^{*2} - 2|a_0|^2 |b_0|^2)$ (21b)

Obviously, the curve obtained from the intersection of these two surfaces determines the evolution trajectory. If the trajectory is closed the system will be stable; on the other hand if the trajectory opens to infinity the system will be unstable. Figure 3 illustrates these two surfaces in three different regimes. Fig. 3(a) shows the stable case, Fig. 3 (b) shows the threshold of instability, while Fig. 3(c) corresponds to an unstable case. In each case the right hand side panel shows the cross section in the $S_0 S_3$ plane. According to this figure, to have a stable solution, the two surfaces described in Eqs. (21) should intersect in the $S_0 S_3$ plane ($S_1 = 0$). In other words, the pair of equations $S_0^2 = S_3^2 + D$ and $S_0 = \frac{\rho}{4\kappa} S_3^2 + C$ should have a valid solution. This means that the combination of the two equations, i.e., $S_0^2 - \frac{4\kappa}{\rho} S_0 + \left(\frac{4\kappa}{\rho} C - D\right) = 0$ should have real solutions. Therefore $\Delta = \left(\frac{4\kappa}{\rho}\right)^2 - 4\left(\frac{4\kappa}{\rho} C - D\right)$ should be a positive quantity. After writing $C$ and $D$ in terms of the initial conditions (Eqs. (21)) the latter condition can be simplified as follows:

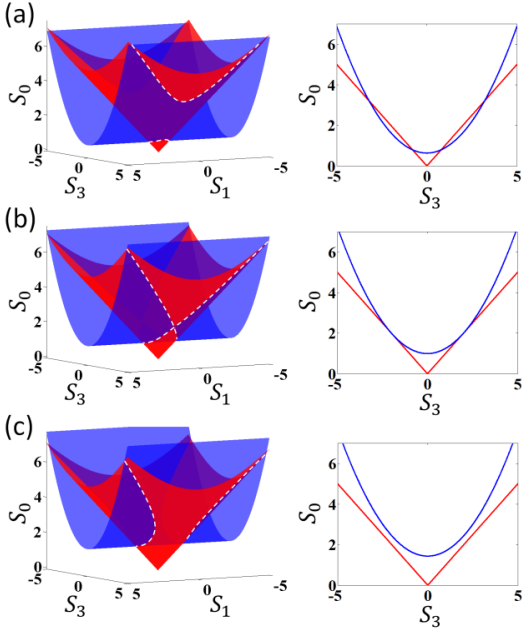

FIG. 3. (Color online) Intersection of the two surfaces described by $S_0^2 = S_1^2 + S_3^2 + D$ (red) and $S_0 = \frac{\rho}{4\kappa} S_3^2 + C$ (blue). The dashed white line shows the intersection curves. For all cases $\frac{\rho}{\kappa} = 1$. The initial conditions used are (a) $a_0 = 0.8$, $b_0 = 0$, (b) $a_0 = 1$, $b_0 = 0$ and (c) $a_0 = 1.2$, $b_0 = 0$. In each case the right hand panel shows a cross section of the left panel in the $S_0 S_3$ plane.

$$\Delta = \left(\frac{\kappa}{\rho}\right)^2 - \left(\frac{\kappa}{\rho}\right)(|a_0|^2 + |b_0|^2) + |a_0|^2 |b_0|^2 \quad (22)$$

According to this relation and based on the initial conditions three different regimes can be distinguished: (a) If $|a_0| < \sqrt{\kappa/\rho}$ and $|b_0| < \sqrt{\kappa/\rho}$ the discriminant $\Delta$ is positive and in this case the PT coupler is stable. (b) If $|a_0| = \sqrt{\kappa/\rho}$ or $|b_0| = \sqrt{\kappa/\rho}$ the discriminant $\Delta$ is zero and the PT coupler lies on the threshold of instability. (c) If $|a_0| > \sqrt{\kappa/\rho}$ or $|b_0| > \sqrt{\kappa/\rho}$ then $\Delta$ is negative and the PT coupler is unstable.

## VII. Conclusions

In conclusion we have studied the Schrödinger equation in the presence of a nonlocal nonlinearity which respects PT symmetry. We showed that such equation shows altogether new behavior. In particular it admits both bright and dark solitons at the same time. The experimental realization of such nonlinearities in a continuous system may be a huge challenge. However, discrete systems like a lattice or a coupled waveguide may facilitate such realization. Therefore, we also considered a discretized version of the PT symmetric nonlinear Schrödinger equation. When only two elements were involved, we showed that such system is fully integrable in terms of elliptic functions. Finally, by using Stokes parameters we obtained an analytical expression for the PT-symmetry breaking instability threshold.